\documentclass[twocolumn,showpacs,preprintnumbers,amsmath,amssymb]{revtex4}

\usepackage{color}
\usepackage{graphicx}
\usepackage{dcolumn}
\usepackage{bm}
\usepackage{amsmath}

\newcommand{\bea}{\begin{eqnarray}}
\newcommand{\eea}{\end{eqnarray}}
\newcommand{\beq}{\begin{equation}}
\newcommand{\eeq}{\end{equation}}
\newcommand{\beqa}{\begin{eqnarray}}
\newcommand{\eeqa}{\end{eqnarray}}

\begin{document}

\preprint{DO-TH 09/10}

\title{Collider Signatures of Minimal Flavor Mixing from Stop Decay Length Measurements}

\author{Gudrun Hiller, Jong Soo Kim and Henning Sedello}
 \affiliation{Institut f\"ur Physik, Technische Universit\"at Dortmund,
  D-44221 Dortmund, Germany}


\begin{abstract}
We investigate the prospects to extract supersymmetric couplings from a decay length measurement at the LHC. Specifically, we exploit the opportunity of a light and long-lived stop 
which is pair-produced  through gluinos in association with like-sign top quarks. 
Any observed finite value of the stop decay length strongly supports models in which flavor is broken  in a minimal way solely by the Standard Model Yukawa couplings.
We find that a 1 picosecond stop lifetime, dominated by $\tilde t \to c \chi^0$ decays, yields macroscopic transverse impact parameters of about 180 microns. If the lightest neutralino is predominantly higgsino or very close in mass to the light stop, the stop lifetime 
even increases and allows to observe stop tracks and
possibly secondary vertices directly. Measuring squark flavor violation with the stop decay length works also with a gravitino LSP if the neutralino is the NLSP. For this case, opportunities from $\tilde t \to c \chi^0 \to c \gamma \tilde G$  decays for very light gravitinos with mass $\lesssim$ keV are pointed out.
\end{abstract}

\pacs{14.80.Ly,12.60.Jv,12.90.+b}
\maketitle

\section{\label{sec:introduction} Introduction}

The past decade has brought great advances in the description of quark
flavor violation in the Standard Model (SM) in terms of both
consistency checks and precision.  Together with the constraints from
$K,D$ and $B$ meson studies on flavor changing neutral currents (FCNC),
this has strong implications: The physics at the TeV-scale cannot
contain much more flavor violation than the SM \cite{Grossman:2009dw}.
Intriguing loopholes exist presently, however \cite{Soni:2009fg}.

A generic framework consistent with current flavor observations is
minimal flavor violation (MFV), where the Yukawa matrices are the sole
source of flavor breaking, as in the SM.  In the context of the
minimal supersymmetric standard model (MSSM), MFV predicts highly
degenerate first and second generation squarks and their mixing with
the third generation to be suppressed by small CKM quark mixing angles
$V_{ij}$ of the order $|V_{cb}|, |V_{ts}| \sim 0.04$ (second-third
generation) or smaller $|V_{ub}| \sim 0.004$, $|V_{td}| \sim 0.01$
(first-third generation).

Is it possible to directly measure such small flavor changing
couplings in squark mixing and support MFV?
We pursue here the proposal of Ref.~\cite{Hiller:2008wp} and exploit the
opportunity of a light stop whose decay to top quarks is
kinematically forbidden. If this stop then decays predominantly via
FCNC into charm and the lightest neutralino $\tilde t \to c \chi^0$,
it has a long life because the coupling between stop and charm is
CKM-suppressed as dictated by MFV.  (In this work $\tilde
t$ and $\chi^0$ always denote the lightest stop and neutralino,
respectively.)  Lifetimes of order picoseconds are possible, and with
boost factors $\gamma \beta \sim 1$ \footnote{An on-shell particle with mass $m$ moving with momentum $\vec p$ has energy $E=\sqrt{|\vec p|^2+m^2}$,
$\gamma=E/m$ and $\beta=|\vec p|/E$.}, this leads to a macroscopic decay
length of the order of a few hundred microns.  The goal of this
work is to analyze the prospects to observe this scenario and extract
superpartner MFV couplings at the LHC.

The paper is organized as follows: In Sec.~\ref{sec:MFV} we briefly review the framework
of a light stop in MFV. In Sec.~{\ref{sec:collider}} the collider signatures are worked out.
The lightest supersymmetric particle (LSP) does not need to be the $\chi^0$.
  In Sec.~\ref{sec:gravitino} we consider the case with a gravitino LSP and the stop or the neutralino being the next lightest supersymmetric particle (NLSP).
In Sec.~\ref{sec:summary} we summarize and conclude.

\section{A Light Stop in MFV \label{sec:MFV}}

Our framework is a generic MSSM setting which is MFV and where the
lightest stop decays predominantly to charm.  (Decays to first
generation up quarks are further CKM-suppressed.)
Many of the contemporary TeV-scale models are MFV, such as gauge and
anomaly mediation and hybrids, {\it e.g.,} \cite{Dermisek:2007qi}, or
by construction, the CMSSM and msugra, see, {\it e.g.,} \cite{Martin:1997ns}.

Within MFV, all flavor changing effects with squarks and quarks are
controlled by the quark Yukawas $\lambda_q$ and CKM angles
\cite{D'Ambrosio:2002ex}.  In particular, the $\tilde t -c -\chi^0 $
coupling $Y$ is parametrically suppressed as
\cite{Hiller:2008wp,Hikasa:1987db}
\begin{equation} \label{eq:suppress} Y \propto \lambda_b^2 V_{cb}
  V_{tb}^* \sim 3 \cdot 10^{-5} \tan \beta^2,
\end{equation}
where the higgsino component of the $\chi^0$ receives an additional
suppression by the charm Yukawa $\lambda_c \sim 10^{-2}$.  The precise
value of $Y$ depends on the composition of the stop and the
neutralino and, to some extent, on the flavor diagonal properties of
the supersymmetry (susy) breaking mechanism.  
Hence, even further suppressions  beyond those from flavor  given in Eq.~(\ref{eq:suppress}) can arise   \cite{Hiller:2008wp}.

In models with flavor blind susy breaking
at the scale of mediation $M$, intergenerational squark mixing is radiatively induced at the weak scale $m_Z$ at order $1/(16 \pi^2) \ln( m_Z/M)$ times a flavor factor, which for the
case of stop-scharm mixing is written in 
Eq.~(\ref{eq:suppress}).
Since the coupling $Y$ is normalized to an average squark mass squared which also receives logarithmic corrections, the dependence on the scale $M$ gets partially cancelled. There
is, however, sensitivity to the relative size of gaugino and flavor-universal scalar masses and trilinear terms.

Note that a small $Y$ is very specific to MFV susy. Other MSSM variants give generically larger values of the flavor factors, such as $Y \propto \lambda_c$ (flavor anarchy) or 
$Y \propto V_{cb} \lambda_c$ (alignement \cite{Nir:1993mx}).

Stop decays into tops, which would otherwise be a leading channel, 
can be forbidden by a suitable spectrum
\beq\label{eq:deltam}
\Delta m = m_{\tilde t} -m_{\chi^0} < m_t,
\eeq
where $m_{\tilde t}, m_{\chi^0}$, and $m_t$ denotes the mass of the
lightest stop, the lightest neutralino, and the top quark, respectively.

To also avoid a contamination from four-body decays $\tilde t \to b \ell \nu \chi^0$,
$\Delta m$ needs to be lower than the bound given in Eq.~(\ref{eq:deltam}).
The actual value is model-dependent and typically in the range $[\mbox{few}(5- 10) ]$ GeV.
This choice is also consistent with the constraints from direct searches~\cite{Abazov:2008rc}.
Neglecting the charm mass
in kinematical factors, the \mbox{$\tilde t \to c \chi^0$} decay rate is then given as
\beq\label{eq:decrat}
\Gamma=\frac{m_{\tilde t} Y^2}{16\pi}\left(1-\frac{m_{\chi^0}^2}{m_{\tilde t}^2}\right)^2
\approx\frac{m_{\tilde t} Y^2}{4\pi}\left(\frac{\Delta m}{m_{\tilde t}}\right)^2.
\eeq
It is suppressed by the small coupling and mass splitting.

With $m_{\tilde t} \sim {\cal{O}}(100) \, \mbox{GeV}$ and
Eqs.~(\ref{eq:suppress})-(\ref{eq:decrat}), stop lifetimes in the
picoseconds range arise. Consequently, the  stop hadronizes before decay.

\section{$\tilde t \to c \chi^0$ Collider Signatures \label{sec:collider}}

We work out collider signatures of a light stop decaying to charm and the lightest neutralino
within MFV susy.
We make a few minimal assumptions on  the susy spectrum only:
The lighter stop is light, and its mass is not too far away from the one of the lightest neutralino.
In addition, we require the mass of the gluino $m_{\tilde g}$ to be sufficiently heavy to
allow for decays to $\tilde t \bar t  $ and $ \tilde t^* t  $.  The masses of all
other squarks  except for the lighter stop should be sufficiently above $m_{\tilde g}$ such
that the gluino decays predominantly into these top plus light stop modes.

\subsection{Open stop production}

We consider stop production in association with like-sign tops in $pp$-collisions
\bea
\label{eq:stop-production} pp \to \tilde g \tilde g
\to t t \tilde t^* \tilde t^*, \bar t \bar t \tilde t \tilde t .
\eea
Such processes arise due to the Majorana nature of the gluinos and, unlike the reaction into $t \bar t\tilde t\tilde t^*$-states, have a controlled background \cite{Kraml:2005kb}.  
Direct stop pair-production, $pp \to \tilde t \tilde t^*$, is disfavored despite its
large cross section of the order of several hundred pb
\cite{Beenakker:1997ut} \footnote{We always assume 14 TeV
center-of-mass energy unless otherwise stated.} due to the
difficult stop identification from charm jets plus missing energy, {\it e.g.,} 
\cite{Yang:2000af,Han:2003qe}. See, however,
\cite{Han:2003qe} for cases if one of the stops decays differently than via FCNCs, 
and \cite{Carena:2008mj}.
We will come back to di-stop production  in Sec.~\ref{sec:photonE}.

For the numerical analysis we use MadGraph/MadEvent 4.4.23 \cite{Stelzer:1994ta} 
and independently our own leading order event
generator with CTEQ6L1 parton distribution functions
\cite{Pumplin:2002vw}. We use for the factorization and renormalization
  scales $\mu=m_{\tilde g}$. With $m_{\tilde t} \sim
{\cal{O}}(100) \, \mbox{GeV}$ and $m_{\tilde g} =500$ GeV
cross sections of 5 pb for the processes
Eq.~(\ref{eq:stop-production}) at the LHC are obtained. 
The cross section decreases to 0.2 pb for $m_{\tilde  g}=1000$ GeV. 
The dependence on the stop mass is very
mild for light stops below the top.  Next-to-leading order corrections yield $K$-factors of 
$\sim 1.5-2$ for di-gluino production at the LHC \cite{Beenakker:1996ch}, suggesting similar
enhancements for top-associated stop production Eq.~(\ref{eq:stop-production}). 
For a lower proton-proton center-of-mass energy of 10 TeV, the cross sections  for 
Eq.~(\ref{eq:stop-production}) are reduced by a factor of  $\sim 0.15$ with respect to those in the 14 TeV case.

\subsection{Stop boost and decay length \label{sec:boost}}

The stop decay lengths $d_i, i=1,2$ are related to the fundamental parameters 
on average \footnote{In our simulations we let the stops decay at the time $t$ with the probability $\Gamma \exp(-t \Gamma$). We use "lifetime" synonymous for the mean lifetime $\tau=1/\Gamma$ throughout this work.}
as
\beq \label{eq:di}
d_i =\frac{(\gamma \beta)_i}{\Gamma} \approx  
\frac{4 \pi p_i}{(\Delta m Y)^2} ,
\eeq
where the $p_i$ denote the magnitudes of the three momenta of the two stops in the
laboratory frame.  In the second step we used  the small mass splitting approximation  $\Delta m \ll m_{\tilde t}$,  see Eq.~(\ref{eq:decrat}). 

A measurement of the decay length can be used to extract $\Delta m Y$ if the stop kinematics
is known and to extract $Y$ if the mass splitting is known also. 
As already stressed, the exact value of $Y$, within MFV, could tell us something about the composition of the lightest stop and neutralino and the susy breaking mechanism.

For $m_{\tilde t} \sim {\cal{O}}(100)\,  \mbox{GeV}$ and
$m_{\tilde g}\sim (500-1000)$ GeV
the stops produced at the LHC through Eq.~(\ref{eq:stop-production})
are typically boosted with $\gamma \beta \sim {\cal{O}}(1-10)$, supporting macroscopic
decay lengths, see Eq.~(\ref{eq:decrat}), as
\beq\label{eq:distance}
d_i \sim 0.5{\rm mm}\ \frac{(\gamma \beta)_i}{5}  \left(\frac{100 \mbox{GeV}}{m_{\tilde t}}\right) \! 
\left(\frac{0.05}{\Delta m/m_{\tilde t}}\right)^2 \!  \! \left(\frac{10^{-5}}{Y}\right)^2 \! \!.
\eeq
The boost factors increase for heavier gluinos, however, at the price of fewer events.

\begin{figure}
  \centering
  \includegraphics[width=0.45\textwidth]{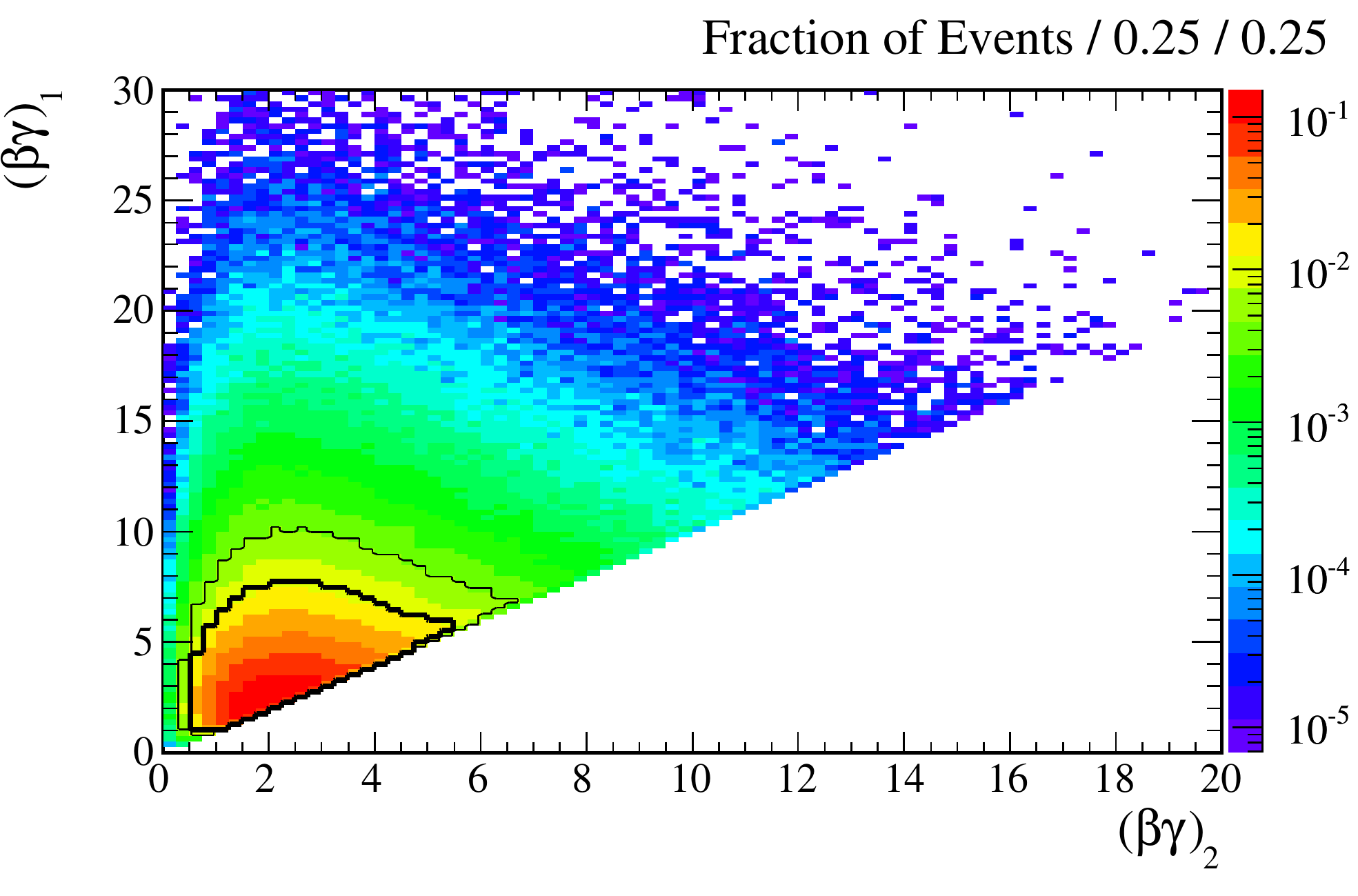}
  \caption{The stop boost factors at the LHC in processes
    Eq.~(\ref{eq:stop-production}) for $m_{\tilde t}=100$ GeV and
    $m_{\tilde g}=500$ GeV. 
      The thick (thin) contour contains $80$\% ($90$\%) of the events.} 
  \label{fig:boost}
\end{figure}
\begin{figure}
  \centering
  \includegraphics[width=0.45\textwidth]{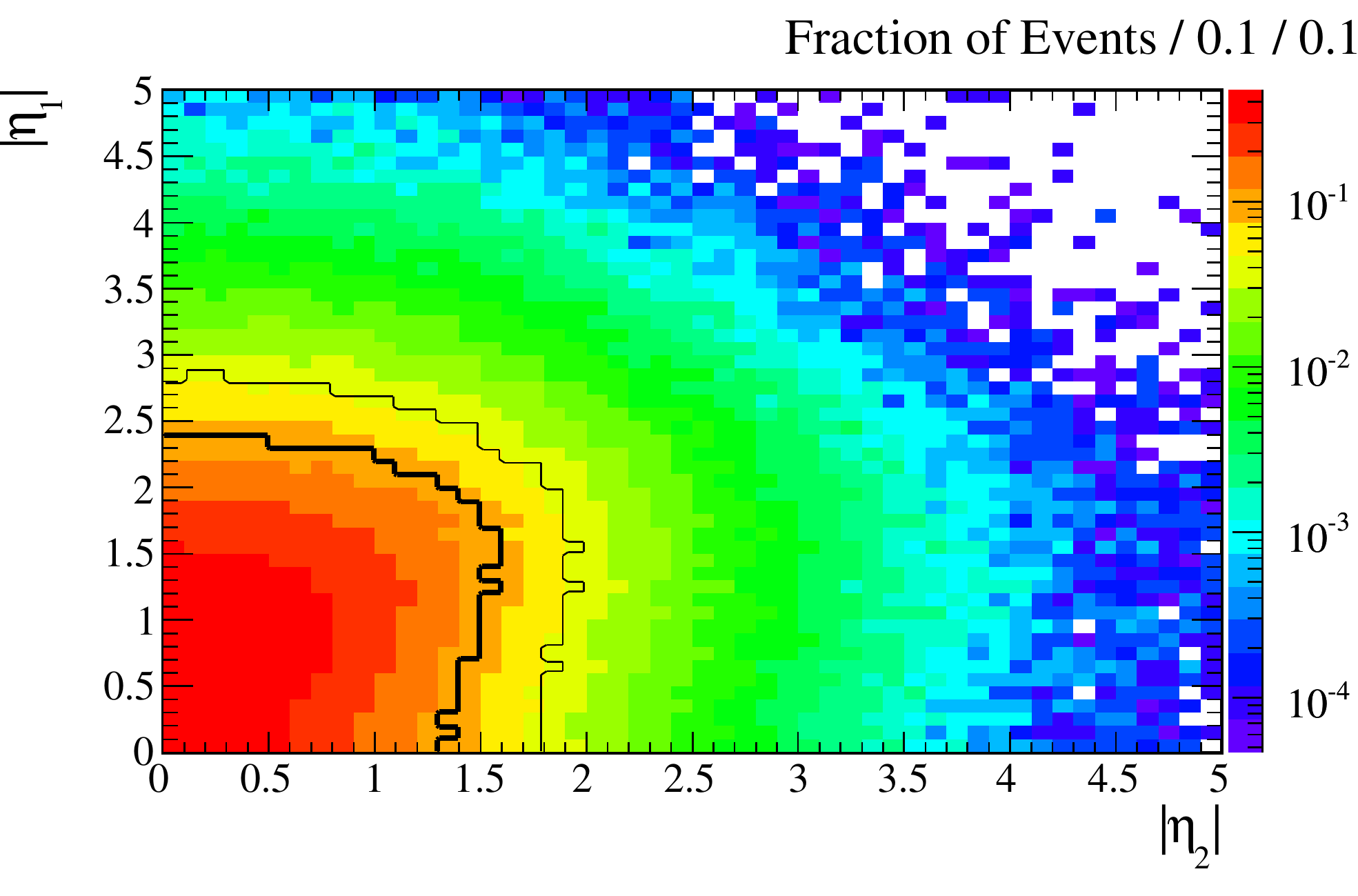}
  \caption{The  pseudo-rapidity distribution of the two stops at the LHC in
    processes Eq.~(\ref{eq:stop-production}) for $m_{\tilde t}=100$
    GeV and $m_{\tilde g}=500$ GeV.  In all shown events
    both stops satisfy $\gamma \beta >1$.  
 The thick (thin) contour contains $80$\% ($90$\%) of the events shown. }
  \label{fig:stop-eta}
\end{figure}
\begin{figure}
  \centering
  \includegraphics[width=0.45\textwidth]{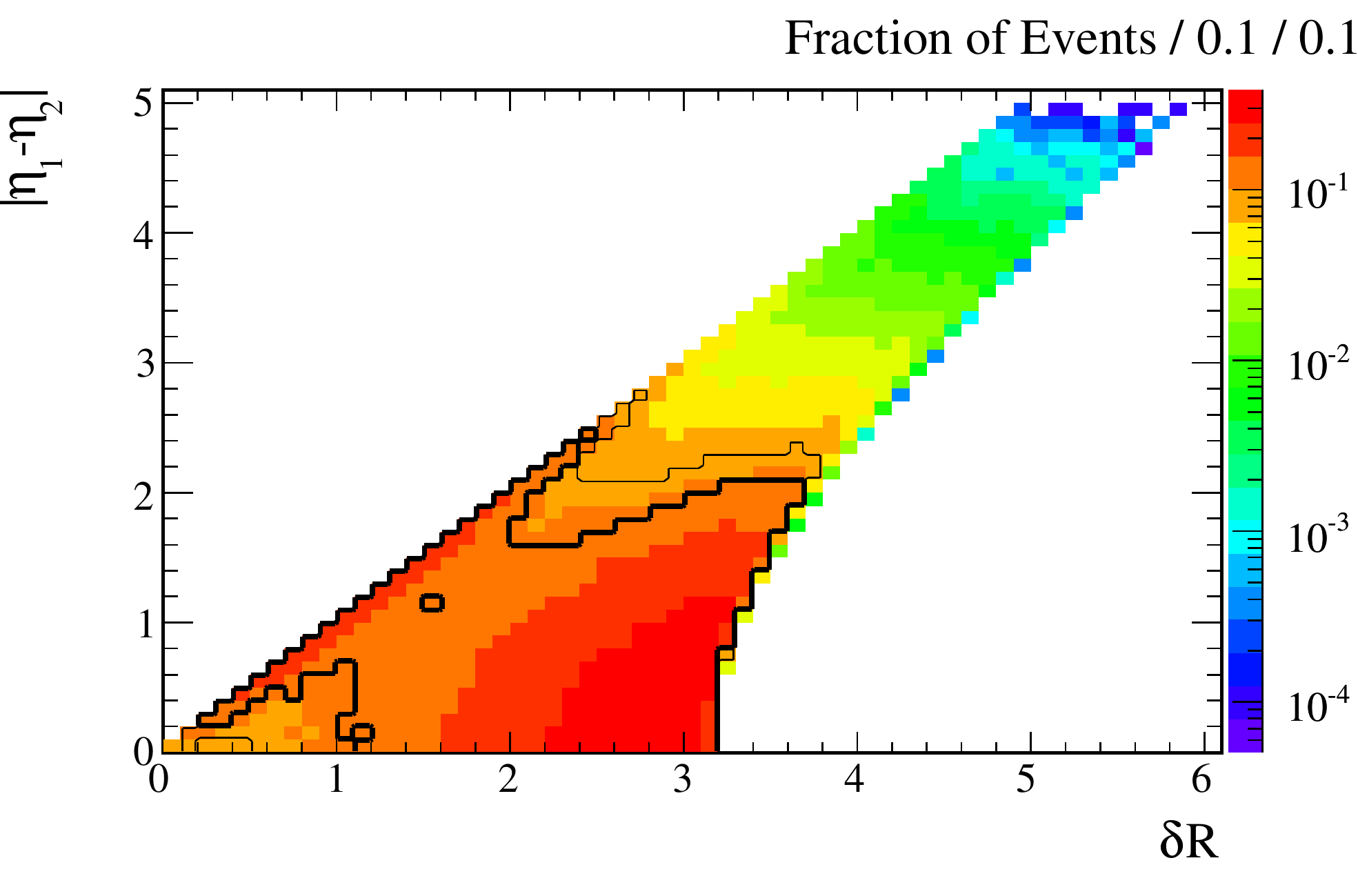}
  \caption{The transverse separation $|\eta_1-\eta_2|$
    versus $\delta R$ of the stops at the LHC in processes
    Eq.~(\ref{eq:stop-production}) for $m_{\tilde t}=100$ GeV and
    $m_{\tilde g}=500$ GeV.  In all shown events  the stops satisfy  $\gamma \beta >1$
    and $|\eta| < 2.5$. The thick (thin)
      contour contains $80$\% ($90$\%) of the events shown.}
  \label{fig:stop-R}
\end{figure}

We show the distribution  of the boost factors of the two stops in
Fig.~\ref{fig:boost} for $m_{\tilde g}=500$ GeV.
For the dominant part of the events, both stops have a significant boost:
$92 $\% of  the events have both $(\gamma \beta)_i >1$
and are produced in the central region, see
Fig.~\ref{fig:stop-eta}. $85 $\% of the events  with both $(\gamma \beta)_i >1$
have $|\eta_i| \leq 2.5$ ($\eta$ denotes the pseudo-rapidity).
As shown in Fig.~\ref{fig:stop-R}, the stops are also well separated in terms of
$|\eta_1-\eta_2|$ or
\beq
\delta R=\sqrt{(\eta_1-\eta_2)^2+(\phi_1-\phi_2)^2} , ~~|\phi_1-\phi_2| \leq \pi ,
\eeq
where $\phi_i$ denote the azimutal angles of the stops in the laboratory frame.
In Figs.~\ref{fig:boost}-\ref{fig:stop-R} the stops in each event are labeled according to $(\beta\gamma)_1\ge(\beta\gamma)_2$.

In a typical scenario the stops will travel distances up to
millimeters, see Eq.~(\ref{eq:distance}). Hence, the stops will decay
before they reach sensible detector material, being away a few cm from
the interaction point within the beam pipe \cite{:2008zzm,:2008zzk}.
Consequently, we have direct access only to the impact parameters
in the transverse plane, $b_i$, for both stops, obtained from
extrapolating the charm trajectories and taking their closest distance
with respect to the beam axis. (In our analysis, we neglect
bending effects due to magnetic fields.)

The impact parameters constitute lower bounds on the decay lengths,
$d_i \geq b_i$.  Already an observation of any finite value of $b_i$
strongly supports MFV since other flavor implementations of the MSSM
generically yield a promptly decaying stop, see Sec.~\ref{sec:MFV}.

Even assuming the optimal situation where all relevant masses are
known, due to the missing energies, the momenta of the stops -- and
the individual lifetimes, see Eq.~(\ref{eq:di}) -- cannot be
reconstructed on an event-by-event basis.  We can however obtain
information about the stop lifetime from the distribution and moments
of impact parameters of the stops.

\begin{figure}
  \centering
  \includegraphics[width=0.45\textwidth]{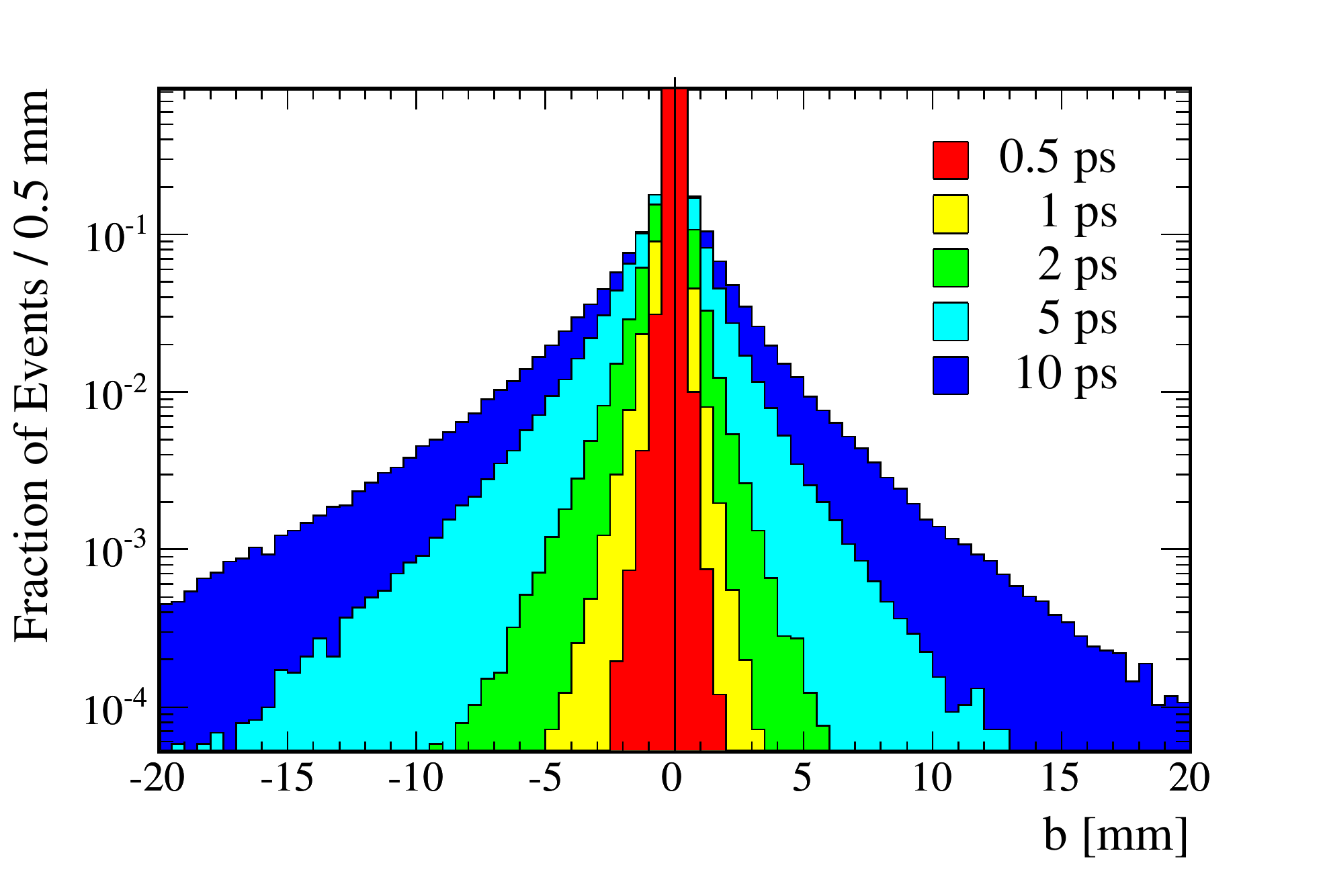}
  \caption{The distribution of impact parameters $b_i$ in mm at the LHC in 
  processes Eq.~(\ref{eq:stop-production}) for different stop lifetimes. The curves with $b_i >0 (<0)$ refer to the stop in each event with larger (smaller) charm $p_T$.}
  \label{fig:bienergy}
\end{figure}

We show the $b_i$ distribution in Fig.~\ref{fig:bienergy} for different values of the stop lifetime 
$\tau =0.5,1,2,5$, and $10 \,\rm{ps}$, corresponding to different values of $Y$, with all input
masses fixed to $m_{\tilde t}=100$ GeV and $m_{\tilde g}=500$ GeV. For a fixed stop lifetime, the dependence of the impact parameter on the mass splitting is negligible as long as it is sufficiently above the charm mass (we used $\Delta m=5$~GeV for the plot). The reason is that for a 
relativistic charm quark the angle between the stop and charm momenta in the laboratory frame becomes insensitive to the charm momentum. 
The two $b_i$ per event are sorted according to the charm transverse momenta $p_T$: positive (negative) values refer to the charm quark with larger (smaller) $p_T$.  
The simulated events show that the distribution is indicative for the stop lifetime.  

Numerically, the average impact parameter can be well approximated as
\beq \label{eq:bsamesing}
\langle b \rangle \simeq 180 \, {\rm \mu m} \cdot \left(  \frac{\tau}{\rm ps}  \right).
\eeq
The formula is largely insensitive to the stop boost factors, {\it i.e.,} the gluino and stop mass
\footnote{For relativistic particles the effect of a boost-enlarged decay length
is compensated by the boost-suppression of the angle between the particle and its decay products,
and vice versa.}. The asymmetry between the two impact parameters
$(\langle b_1 \rangle -\langle b_2 \rangle )/(\langle b_1 \rangle +\langle b_2 \rangle)$ is $24 \%$.

If the stop coupling $Y$ is even further suppressed than in
Eq.~(\ref{eq:suppress}), for instance due to small wino/bino
components in the $\chi^0$, the decay lengths may be measured
directly. Specifically, inspecting Eq.~(\ref{eq:distance}) for
\beq
\frac{\Delta m}{m_{\tilde t}} Y \lesssim 5 \cdot 10^{-8},
\eeq
decay lengths of a few centimeters and larger arise; hence, the
(hadronized) stops
produce tracks from interactions with the detector material, see
\cite{Kraan:2004tz}. 
The appearance of stop tracks  is advantageous because more
information can be obtained without relying solely on the charm jets.

Very small values of
\beq
\frac{\Delta m}{m_{\tilde t}}Y \lesssim 4 \cdot 10^{-9}
\eeq
yield stops which travel the detector undecayed. Searches from the CDF
experiment at the Tevatron have ruled out this scenario for light
stops with masses below 249~GeV~\cite{Aaltonen:2009kea}.

\subsection{Stop events}

Stops decaying to charm jets and missing energy in association with 
like-sign tops, Eq.~(\ref{eq:stop-production}), can be searched for in the final states
\bea
 bb jj \ell^+ \ell^+ \not \! \! E_T, ~~ \bar b \bar b jj \ell^- \ell^- \not \!  \!E_T , ~~~~ \ell=e, \mu,
 \eea
 requiring leptonically decaying top quarks.
As discussed in \cite{Kraml:2005kb}, signals can be separated from the
SM and susy backgrounds
at the LHC by employing  cuts on same-sign isolated lepton pairs, missing transverse
energy, $\not\!\!\!E_T$, and the transverse momenta of the four ($c$ and $b$) hardest jets.
In our simulation $\not\!\!E_T$ consists of the two lightest neutralinos and
the two neutrinos from the leptonic decays of the top quarks.

\begin{figure}[h]
  \centering
  \includegraphics[width=0.45\textwidth]{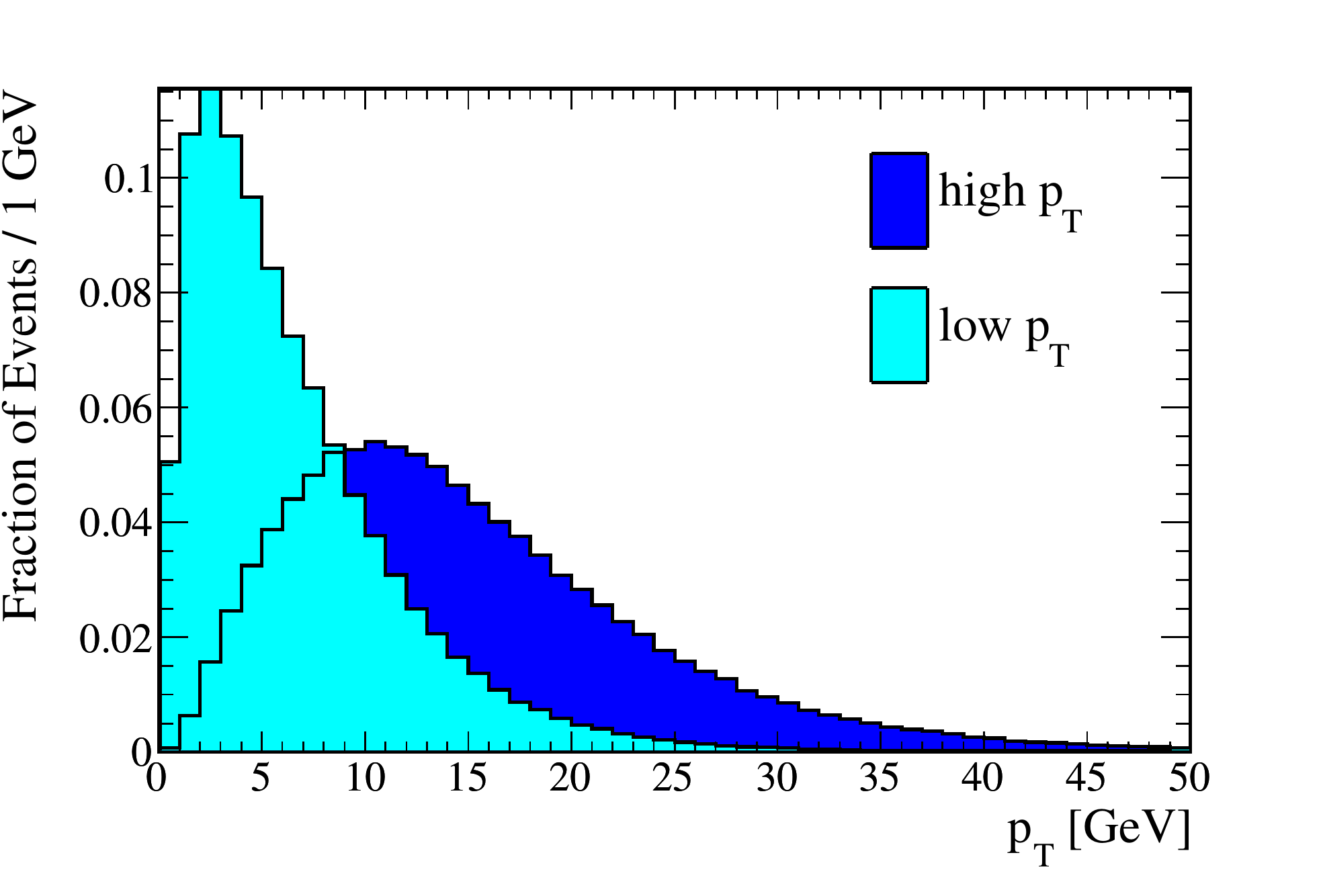}
  \caption{The $p_T$ distributions of the charm quarks stemming from
    the decays of light stops produced in processes
    Eq.~(\ref{eq:stop-production}) for $m_{\tilde t}=100$ GeV,
    $m_{\tilde g}=500$ and $\Delta m=5$ GeV. The light blue (lighter shaded)
    histogram corresponds to the charm quarks with per event lower  $p_T$.}
  \label{fig:charm-e}
\end{figure}

\begin{figure}[h]
  \centering
  \includegraphics[width=0.45\textwidth]{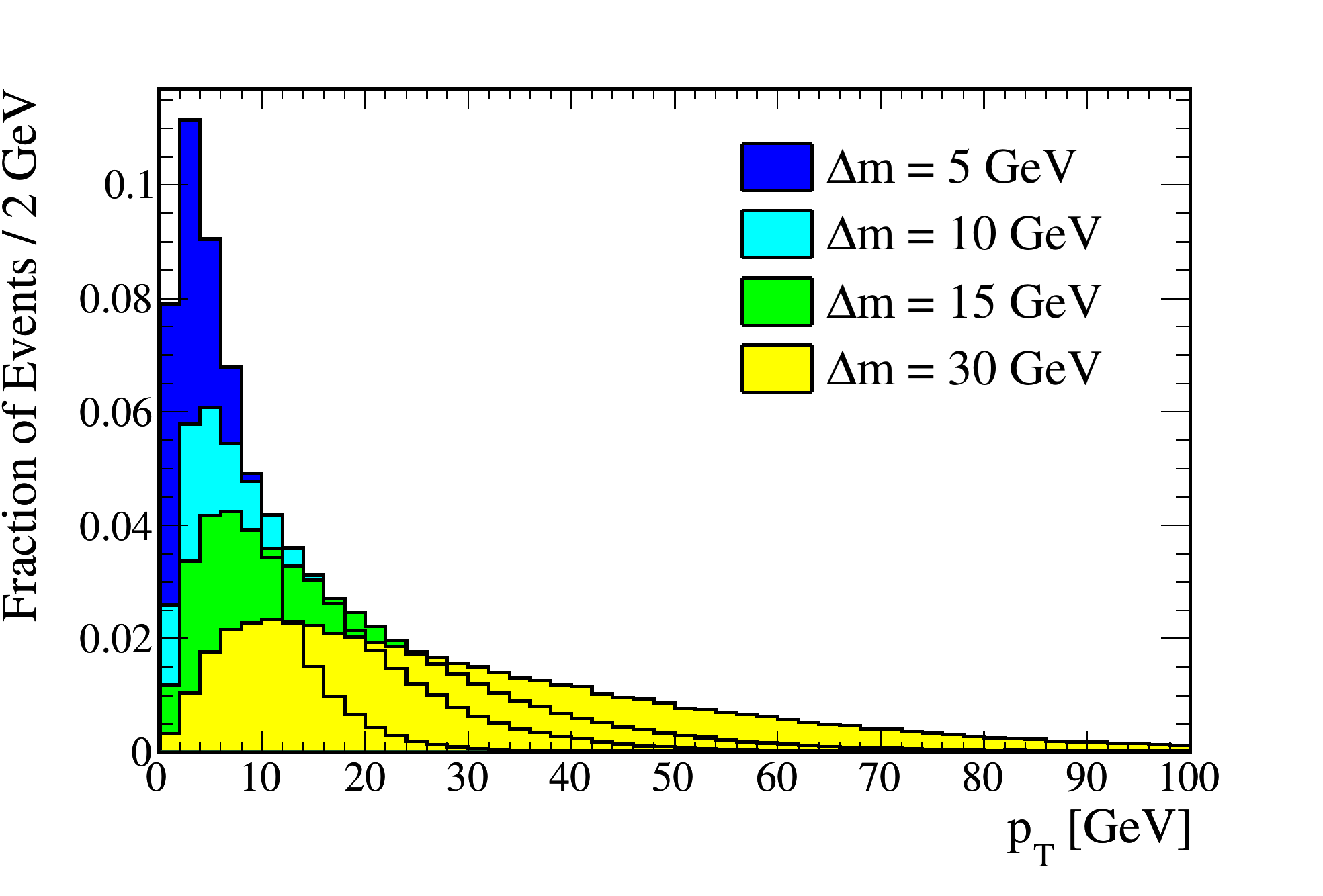}
  \caption{The $p_T$ distribution of the lower $p_T$ charm quarks stemming
    from the decays of light stops produced in processes
    Eq.~(\ref{eq:stop-production}) for $m_{\tilde t}=100$ GeV,
    $m_{\tilde g}=500$ GeV, and different mass splittings $\Delta m$.}
  \label{fig:charm-massgap}
\end{figure}

In Fig.~\ref{fig:charm-e} we show the $p_T$ distributions of the two charm quarks
for $m_{\tilde t}=100$ GeV, $m_{\tilde g}=500$ GeV, and $\Delta m=5$ GeV. 
The charm $p_T$ increases with increasing stop boost, hence, for heavier gluinos
and lighter stops.
A small stop-neutralino mass splitting supports a stop branching ratio dominated by
FCNC, ${\cal{B}}(\tilde t \to c \chi^0) \simeq1$,
and a long stop lifetime, see Eq.~\eqref{eq:decrat},  but it also makes the charm jets less energetic.
The $p_T$ distribution of the charm quarks with lower $p_T$ per event is shown for different values of $\Delta m$ in Fig.~\ref{fig:charm-massgap}. We kept the stop mass fixed, hence,
lowered the mass of the lightest neutralino correspondingly.
The fraction of events surviving various  $p_T$ cuts for different  mass splittings is given in
Tab.~\ref{tab:charm-cuts}.

\begin{table}[h]
  \centering
  \begin{tabular}{c|rrrr}
    $p_T^{min}$    & $30$ GeV & $35$ GeV & $40$ GeV & $50$ GeV\\
    \hline\hline
    $\Delta m = 5$ GeV   & $0.4$\% &	$0.2$\% &	$0.08$\% & $0.02$\%\\
    $\Delta m = 10$ GeV   & $7$\% &	$4$\% & 	$2$\% &	   $1$\%\\
    $\Delta m = 15$ GeV   & $18$\% &	$13$\% &	$9$\% &	   $4$\%\\
    $\Delta m = 30$ GeV   & $45$\% &	$38$\% &	$32$\% &   $22$\%\\
  \end{tabular}
  \caption{The fraction of events of  processes
    Eq.~(\ref{eq:stop-production}) with both charm quarks surviving a $p_T$ cut for different
    values of $p_T^{min}$ and  $\Delta m$ for $m_{\tilde t}=100$ GeV and
    $m_{\tilde g}=500$ GeV.}
  \label{tab:charm-cuts}
\end{table}

With a factor of $(2/9)^2$ for leptonically  ($\mu$ or $e$) decaying top quarks, 
the expected number of events for an integrated luminosity of $3 (10) \,\mbox{fb}^{-1}$ and
$\sigma_{NLO} =7.5$ pb (for
 $m_{\tilde t}=100$ GeV and $m_{\tilde g}=500$ GeV) is  about
$1110(3700)$ times loss due to kinematical cuts and detector effects.
{}From the $p_T$ cuts of $20 \, {\rm GeV}$ for the charged leptons, $50 \, {\rm GeV}$ for each $b$, and $100 \, {\rm GeV}$ for $\not\!\!E_T$\cite{Kraml:2005kb}, a reduction factor of 0.42 arises. 
The latter drops to 0.08 with a $b$-tagging efficiency of
$43\%$. Note that our analysis does not employ charm-tagging.

The $p_T$ cuts on the charm quarks  have a most significant impact on the event rate, see Tab.~\ref{tab:charm-cuts}.
The number of reconstructed stops with a long life then depends strongly on the stop-neutralino
mass splitting.
Assembling all factors and neglecting further detector effects, up to 
$\sim 100$ events can be expected for $10 \,\mbox{fb}^{-1}$ at our benchmark point at the LHC.

\section{Long-lived stop and gravitino LSP \label{sec:gravitino}}

So far we did not make use of an LSP-feature of the lightest neutralino. We only assumed tacitly that the $\chi^0$ is stable on the size of the detector. We ask now about the implications for the stop decay length analysis outlined in the previous section  if the gravitino  $\tilde G$ is the LSP. 

Interactions of the gravitino with the other MSSM particles are down 
by the reduced Planck mass  $m_{\rm Pl}=1/\sqrt{8 \pi G_N} =2.4 \cdot
10^{18} \, \rm{GeV}$ and typically yield very slow decays. This suppression can, however,  be lifted 
with small masses $m_{3/2}$ of the gravitino due to its goldstino component. Indeed, gravitino masses are linked to the susy breaking $F$-terms, $m_{3/2}=F/(\sqrt{3} m_{\rm Pl})$, and 
 can be very low, even below eV, depending on the mediation mechanism, see
 {\it e.g.,} \cite{Martin:1997ns} for a brief overview and references therein.

We consider two cases:  In the first one, discussed in Sec.~\ref{sec:stopnlsp}, the lighter stop is the NLSP. In the second scenario, addressed in Sec.~\ref{sec:neutralino}, we assume 
$m_{\tilde t} > m_{\chi^0} > m_{3/2}$ 
such that the stop can decay to both the gravitino and the lightest neutralino, which is the NLSP. 
The full decay chain $\tilde t \to c \chi^0 \to c \gamma \tilde G$ arising in the second case is discussed in Sec.~\ref{sec:photonE}.

We assume that the decay $\tilde t \to t \tilde G$  is kinematically closed.

 \subsection{Stop NLSP \label{sec:stopnlsp}}
 
If the lightest stop is the NLSP, it decays directly via FCNC to the gravitino. 
The decay rate can be written as
\beq \label{eq:stop-grav}
\Gamma(\tilde t  \to c  \tilde G) = \frac{ z^2}{48 \pi} \frac{m_{\tilde t}^5 }{ m_{3/2}^2 m_{\rm Pl}^2} 
\left( 1-\frac{m_{3/2}^2}{m_{\tilde t}^2}\right)^4 ,
\eeq
where $z=\sqrt{|z_L|^2 +|z_R|^2}$, and $z_L(z_R)$ denotes the coupling to left (right)-chiral
charm quarks. Here we neglected the mass of the charm quark in the phase space calculation.
Within MFV, the $z_{L,R}$-couplings are CKM- and Yukawa-suppressed as
\beq \label{eq:zLR}
z_L \propto \lambda_b^2 V_{cb} V_{tb}^* , ~~~~
z_R \sim \lambda_c z_L  \ll z_L .
\eeq
 Evaluating Eqs.~(\ref{eq:stop-grav})-(\ref{eq:zLR}) for a gravitino with mass not much lower than the one of the lighter stop to ensure dominance of the FCNC over the 
 tree level four-body $\tilde t \to b \ell \nu \tilde G$ decays,
 one finds that the stop always leaves the detector undecayed, {\it i.e.,} the proper decay length is larger than $\sim 10$ m. Again, this is ruled out for light stops with mass below
 249 GeV by CDF \cite{Aaltonen:2009kea}. 
 
We conclude that in the light stop NLSP scenario with a gravitino LSP, there will not be
 a decay length measurement indicative for the flavor properties of the susy breaking from charm plus missing energy inside the detector.

\subsection{Neutralino NLSP \label{sec:neutralino}}

In the presence of a gravitino with mass below $m_{\chi^0}$, the lightest neutralino
can decay via $\chi^0 \to X \tilde G$, where X can be the photon or the $Z$-boson or a neutral Higgs. While the estimation of the exact NLSP lifetime depends on its composition and, if  higgsino
components are present, on the presently unknown Higgs masses and mixing angles, we ask here under which circumstances the $\chi^0$ decays inside of the detector.

Assuming a significant bino/wino fraction in the $\chi^0$, the gravitino decay rate can be written as \cite{Martin:1997ns} 
\beq
\Gamma(\chi^0 \to \gamma \tilde G) \simeq \frac{1}{48 \pi} \frac{m_{\chi^0}^5}{ m_{3/2}^2 
m_{\rm Pl}^2}
+{\cal{O}}\left( \frac{m_{3/2}^2}{m_{\chi^0}^2}\right) ,
\eeq
yielding lifetimes for the lightest neutralino as
\beq \label{eq:chilife}
\tau_{\chi^0} \simeq 10^{-6} s \left( \frac{100 \, {\rm GeV}}{m_{\chi^0}} \right)^5
\left( \frac{m_{3/2}}{4 \, {\rm keV}} \right)^2 .
\eeq
This constitutes  a lower bound on $\tau_{\chi^0} $ because the decay rates to
the massive bosons are further phase space suppressed.

\begin{figure}
  \centering
  \includegraphics[width=0.45\textwidth]{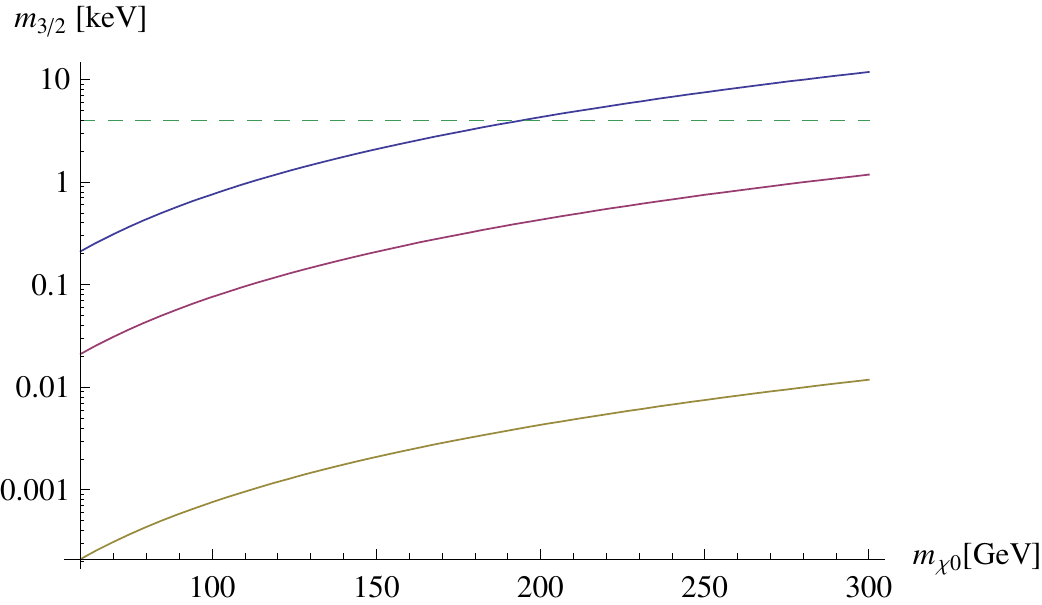}
  \caption{Contours of fixed $\chi^0$ lifetimes in the
$m_{\chi^0}$-$m_{3/2}$ plane as in 
Eq.~(\ref{eq:chilife}). The solid curves correspond, from top to bottom, to
$c \tau_{\chi^0}=10 {\rm m}, 10 {\rm cm}, 10 {\rm \mu m}$. The dashed
line indicates the warm dark matter constraint $m_{3/2} > 4 \, {\rm keV}$ \cite{Viel:2007mv}.}
  \label{fig:mGmchi}
\end{figure}

In Fig.~\ref{fig:mGmchi} we show contours of fixed $\chi^0$ lifetimes in 
the $m_{\chi^0}$-$m_{3/2}$ plane calculated according to Eq.~(\ref{eq:chilife}). The three solid curves (from top to bottom) indicate the regions
below which the $\chi^0$ decays inside of the detector ($c \tau_{\chi^0}= 10 {\rm m}$),
above which a secondary vertex may be observed directly ($c \tau_{\chi^0}= 10 {\rm cm}$),
and below which the decay appears to be prompt ($c \tau_{\chi^0}= 10 {\rm \mu m}$).
Here we ignored effects from the parent stop decay length, which push the solid
curves to lower gravitino masses. 
Higgsino admixture and subdominance of decays to photons, hence a longer $\chi^0$ lifetime, will shift all solid curves downward as well.
Also shown in Fig.~\ref{fig:mGmchi} is the lower bound on the gravitino mass 
$m_{3/2} >4 \, {\rm keV}$ obtained from assuming that the gravitino is responsible for the
observed warm dark matter density in the universe \cite{Viel:2007mv} (dashed line).

The lighter stop can also decay directly to the gravitino.  We use
$\tilde t \to b W \tilde G$ with subsequent decay of the $W$-boson
into two light SM fermions as an estimate for multi-body final states.
In the limit where the charginos and all squarks except for the NLSP
are heavy and the gravitino is light, the decay rate for $\tilde t \to
b W \tilde G$ can be written as \cite{Sarid:1999zx}, within MFV,
\begin{multline}
  \label{eq:stop-grav-W} \Gamma(\tilde t \to b W \tilde G) = \frac{
  \alpha |V_{tb}|^2 }{384 \pi^2 \sin^2\! \theta_W } \frac{m_{\tilde
    t}^5 }{ m_{3/2}^2 m_{\rm Pl}^2}\\
\times\left[|c_L^2\,|I\Big(\frac{m_W^2}{m_{\tilde t}^2},\frac{m_t^2}{m_{\tilde
      t}^2}\Big)
  +|c_R^2|\,J\Big(\frac{m_W^2}{m_{\tilde t}^2},\frac{m_t^2}{m_{\tilde t}^2}\Big)\right].
\end{multline}
Here $m_W$ denotes the mass of the $W$-boson, and 
$c_L$($c_R$) parametrizes the amount of the lighter stop's 
$\tilde t_L$($\tilde t_R$)-content, 
 {\it e.g.,} \cite{Martin:1997ns},
which is not fixed by requiring MFV.  Up to small effects from 
intergenerational squark mixing is $|c_R^2|+|c_L^2|=1$.
The phase space functions $I$ and $J$ are given in~\cite{Sarid:1999zx}.

Comparing Eq.~(\ref{eq:stop-grav-W}) to Eq.~(\ref{eq:decrat}), one
finds that for
\beq \label{eq:m32low}
m_{3/2} \gtrsim \left( \!\!\begin{array}{c} 0.3\\ 24 \end{array}
  \!\!\right) {\rm meV} \left( \frac{5 \cdot 10^{-7}}{ Y \Delta
    m/m_{\tilde t}} \right) \!,~ m_{\tilde
  t}=\left(\!\! \begin{array}{c} 100\\ 150 \end{array}\!\! \right)
{\rm GeV}, \eeq
FCNC stop decays into the lightest neutralino are more rapid than
those directly into the gravitino. The above bound on the gravitino mass
stems from assuming a purely right-handed lightest stop. A left-handed 
$\tilde t$
or left-right admixture allows for  similar but somewhat smaller values of $m_{3/2}$
than in those in Eq.~(\ref{eq:m32low}).

For the large splittings between the mass of the $\tilde t$ and the $\tilde G$
considered here, within MFV, the FCNC decays Eq.~(\ref{eq:stop-grav}),
are suppressed with respect to the ones from the charged current
decays Eq.~(\ref{eq:stop-grav-W}).  Note also that for most of the
parameter space, the lower limit on $m_{3/2}$ in Eq.~(\ref{eq:m32low})
is not more constraining than the requirement of having a not too
light superpartner spectrum.

We conclude that for gravitinos with mass above a few keV, the neutralino NLSP will
decay outside of the detector and that our analysis for $\tilde t \to c \chi^0$ outlined in Sec.~{\ref{sec:collider}} is unaffected.
The case of lighter gravitinos is discussed in the next section.

\subsection{Stop pairs and energetic photon signals \label{sec:photonE}}

Following the analysis of the neutralino lifetime of the previous 
Sec.~\ref{sec:neutralino}, 
a spectrum $m_{\tilde t} > m_{\chi^0} > m_{3/2}$ together with a very light gravitino
$m_{3/2} \lesssim$ keV allows for the exciting 
possibility  that the neutral bosons $X$ from $\tilde  t \to c \chi^0 \to c X \tilde G$ decays are seen
within detectors such as those at the LHC.

We focus here on the case where $X$ is  a photon;
hence, we assume a significant gaugino content in the lightest
neutralino. The photons are energetic, with energies of $\sim m_{\chi^0}/2$ in the rest frame of the
$\chi^0$. (Since $m_{3/2} \ll m_{\chi^0}$, the gravitino can be taken as massless in the kinematics.)

\begin{figure}[h]
  \centering
  \includegraphics[width=0.45\textwidth]{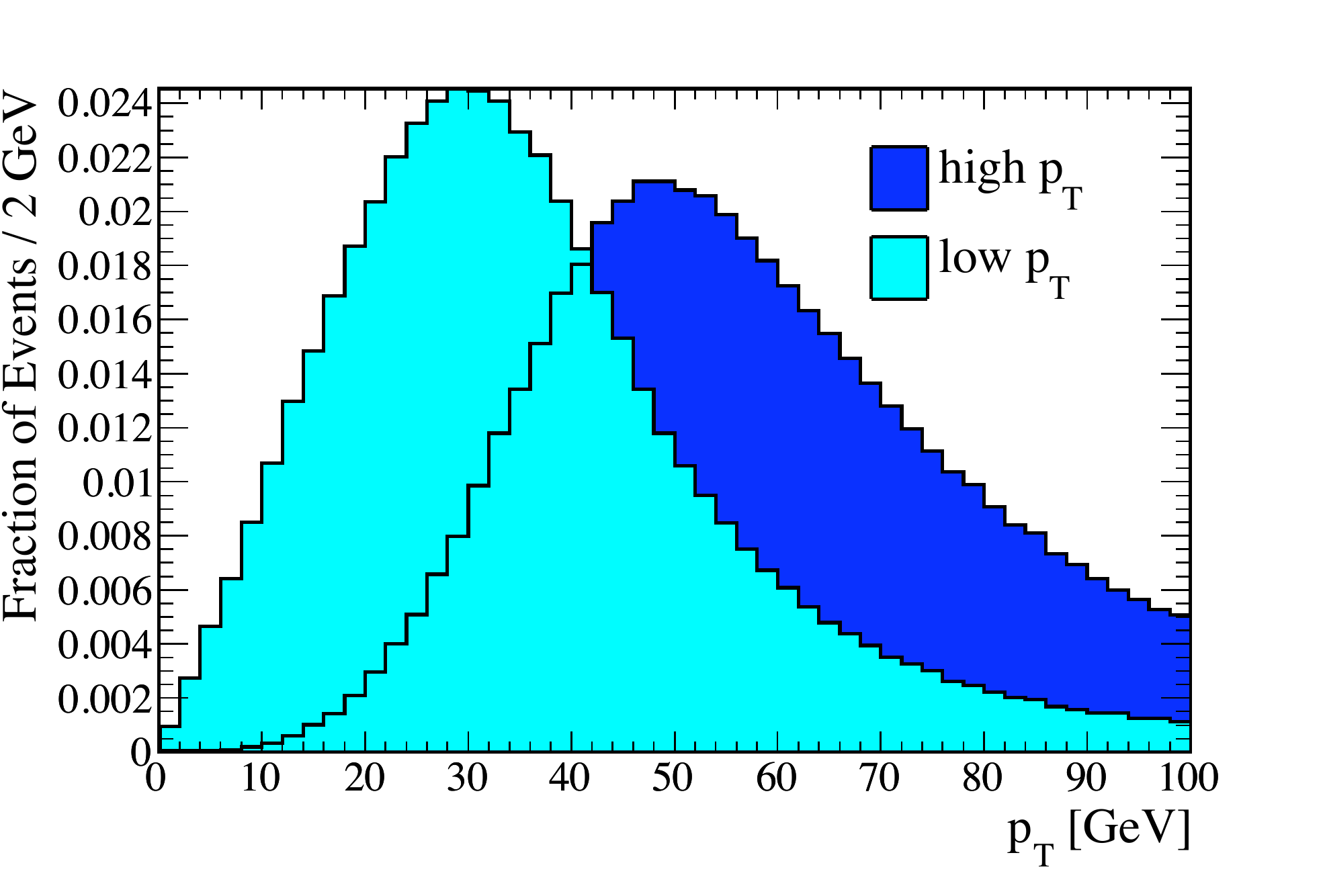}
  \caption{The photon $p_T$ distributions stemming
    from the decays $\tilde t  \to c \chi^0 \to c \gamma \not  \! \! \!E_T$ of
    stops produced in $pp \to \tilde t \tilde t^*$ at the LHC
    for $m_{\tilde t}=100$ GeV. The light blue (lighter shaded)
    histogram corresponds to the photons with per event lower  $p_T$.}
  \label{fig:photons}
\end{figure}

In Fig.~\ref{fig:photons}, we show the transverse momentum spectra of the photons from
$\tilde t  \to c \chi^0 \to c \gamma \tilde G$ with  the stops pair-produced in
$pp \to \tilde t \tilde t^*$ at the LHC \footnote{We use PHYTHIA 6.4.19
  \cite{Sjostrand:2006za} as a framework to calculate the decays of the
  neutralinos.}. We use $m_{\tilde t}=100$ GeV  and ${\cal{B}}(\chi^0 \to \gamma \tilde G)=1$. We also use $\Delta m=5$ GeV, but for $\Delta m \ll m_{\chi^0}$ the
dependence of the photon spectra on the mass splitting is very mild.
Note that constraints on the relevant sparticle masses and compositions
from the recent CDF analysis obtained within gauge mediated susy breaking (GMSB)
\cite{Aaltonen:2009tp} may apply, see \cite{hks2}.

The hard photons are a distinctive signature to efficiently suppress backgrounds 
\cite{hks2,Shirai:2009kn}, and
stop pair-production $pp \to \tilde t  \tilde t^*$ becomes relevant for the LHC. 
While a full simulation of long-lived  stops with energetic photons from NLSP decays is beyond the scope of this work, we briefly give the features of
stop pair-production $pp \to \tilde t  \tilde t^*$ and its impact on stop decay length measurements. 

Compared to  top-associated production worked out in Sec.~\ref{sec:collider}
the boost factors of the stops from di-stop production are somewhat smaller.
Numerically, the average impact parameter can be well approximated as
\beq
\langle b \rangle \simeq 100 \, {\rm \mu m} \cdot \left(  \frac{\tau}{\rm ps}  \right) ,
\eeq
which is roughly half as large as the one in top-associated stop production given in
Eq.~(\ref{eq:bsamesing}) and also macroscopic.
\begin{figure}[h]
  \centering
  \includegraphics[width=0.45\textwidth]{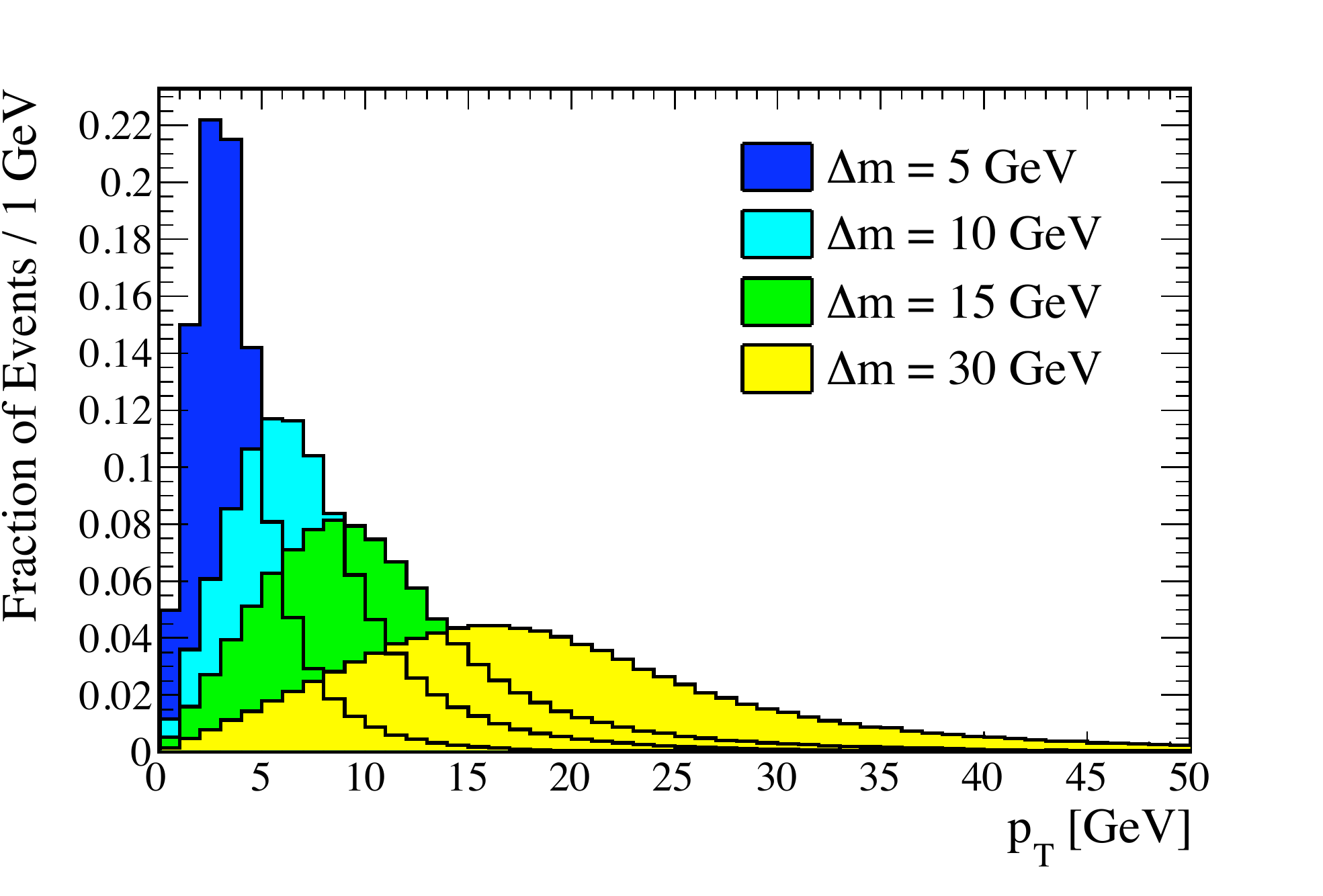}
  \caption{The $p_T$ distribution of the lower $p_T$ charm quarks stemming
    from the decays of light stops produced in $pp \to \tilde t \tilde t^*$ at the LHC
     for $m_{\tilde t}=100$ GeV and different mass splittings $\Delta m$.}
  \label{fig:charm-massgapphoton}
\end{figure}
In Fig.~\ref{fig:charm-massgapphoton} we show the $p_T$ distribution of the 
charm quarks with lower $p_T$ arising in stop decays from $pp \to \tilde t \tilde t^*$.

The advantage of direct stop pair-production over 
same-sign top-associated production
Eq.~(\ref{eq:stop-production}) is a substantially larger production cross section
and the independence of the gluino mass, which might be heavier than a TeV.
Signal loss due to stoponium formation of the di-stops with small relative momentum upon production is about a percent effect  for light stops of ${\cal{O}}(100)$ GeV \cite{Drees:1993uw}.

The photons are  further useful 
for the stop decay length extraction and the determination of mass and mixing parameters. 
Independent of the stop production mechanism,
the following scenarios can arise within susy with MFV:

\begin{itemize}
\item[--] The $\tilde t$ gives a secondary vertex, and the photon points at it.
The prompt NLSP-$\chi^0$ decay yields an upper bound on its lifetime and on the
gravitino mass of the order $m_{3/2} \lesssim {\cal{O}}(10)\, \rm{eV}$, 
see Fig.~\ref{fig:mGmchi}.
The secondary stop decay vertex can be reconstructed using three 
handles: the charm jet, the stop track, and the pointing photon.

\item[--] The $\tilde t$ gives a secondary vertex, and the photon originates from a  tertiary
vertex. The $\chi^0$ decay length is informative of the gravitino mass
and the neutralino mass and
admixture. Approximately, the gravitino mass ranges between
a few $ 0.01\,  \rm{keV}$ and a few keV, see Fig.~\ref{fig:mGmchi}.
The bound on $m_{3/2} $ gets lowered with higgsino admixure in the lightest neutralino
and also with large $\chi^0$ boosts.

\item[--] The $\tilde t$ decays before reaching the innermost  layer of the detector. A finite
impact parameter $b \neq 0$ from charm is observed. The $\chi^0$ decays before reaching detector material. The upper bound on the decay length of the $\chi^0$ implies
$m_{3/2} \lesssim 1 \rm{keV}$, see Fig.~\ref{fig:mGmchi}.

\item[--] The $\tilde t$ decays before reaching the innermost layer of the detector. A finite
impact parameter from charm is observed. The production vertex of the energetic photon is seen.
It can be used to give an upper bound on the lifetime of the $\chi^0$. The
range for the gravitino mass is as in the second scenario.

\end{itemize}

Photons in ATLAS/CMS can be  seen in the calorimeter and as converted photons in the tracker.
LHC studies exist for promptly produced NLSPs decaying to  gravitinos
with non-pointing photons  \cite{Kawagoe:2003jv}  and prompt photons~\cite{Hamaguchi:2008hy}. Also see 
\cite{Carena:2002wz}  for a Tevatron study with promptly decaying stops.

\section{\label{sec:summary} Summary}

After measuring the parameters responsible for quark flavor violation in the SM, it is even more
obvious to ask about the flavor quantum numbers of the SM partners related to electroweak 
symmetry breaking at TeV-energies.
We elaborate here on the prospects of a decay length measurement that probes the
amount of flavor violation of susy breaking couplings in the squark sector at the LHC.

Generically, a light stop is long-lived only if flavor is broken minimally, {\it i.e.}, purely by the quark 
Yukawa matrices. 
(An exception is an unstable stop LSP decaying slowly to SM fermions in models with broken $R$-parity.) While all such MFV models give strongly CKM-suppressed intergenerational stop mixing,
whether one then sees  in the decay $\tilde t \to c \chi^0$ an impact parameter,
$c \tau \gtrsim {\cal{O}}(50 \mu \mbox{m})$,
or a track possibly with secondary vertex, ${\cal{O}}(\mbox{few cm}) \lesssim c \tau \lesssim  10 \mbox{m}$, 
depends on the composition of the stop and its decay products.
If the mass  splittings and mixings are known, the decay length allows further to obtain information
on the mediation mechanism of susy breaking.

Light stops are produced at the LHC with controlled backgrounds through gluinos in association with like-sign tops. We find substantial stop boost factors $\gamma \beta \sim 1-10$ and cross sections in the pb range for gluino masses not exceeding a TeV and stop masses
around ${\cal{O}}(100)$ GeV. Depending strongly on the stop-neutralino mass difference,
up to  $\sim 100$ events can be expected for $10 \,\mbox{fb}^{-1}$ integrated luminosity.

Squark flavor violation  can be probed by the stop decay length with a neutralino or a gravitino LSP as long as in the latter case the neutralino is the NLSP. If the neutralino is not
the dark matter particle, the spectrum can be heavier than ${\cal{O}}(100 \, \rm{GeV})$ 
while keeping the splitting
between the lightest stop and neutralino sufficiently small.

The gravitino LSP provides a further opportunity if the gravitino is very light, below a few keV.
Besides posing no cosmological gravitino problem \cite{Pagels:1981ke}, 
in this case the energetic photon from
$\tilde t \to \chi^0 c \to \gamma  \tilde G c$ can be emitted inside of the detector. This additional
signature is advantageous for both the stop and the neutralino lifetime determination.
It also suppresses backgrounds such that studies based on $pp \to \tilde t \tilde t^*$ with large
stop production cross sections independent of the gluino mass come into reach at the LHC.

We conclude that collider searches for displaced or secondary vertices are a promising area for explorations of many aspects of TeV scale physics.

\begin{acknowledgments}
We are indebted to  Laura Covi, Manuel Drees, David Morrissey, Michael Ratz 
and Carlos Wagner for useful discussions. This work is supported in
part by the Initiative and Networking Fund of the
Helmholtz Association, contract HA-101 ("Physics at the Terascale")
and the German-Israeli-Foundation (G.I.F.).
G.H.~gratefully acknowledges the hospitality and
stimulating atmosphere provided by the Aspen Center for Physics during the final
phase of this work.
\end{acknowledgments}

\end{document}